# CME activities on spotless days during descending phase of solar cycles 23 and 24

Dipali Burud[1], Rajmal Jain[2] , Arun K. Awasthi[3], N. Gopalswamy[4]

(1)Sh. M. M. Patel Institute of Sciences and Research,Kadi Sarva Vishwavidyalaya, Gandhinagar, India

(2) IPS Academy, Rajinder Nagar, AB ROAD, Indore, Madhya Pradesh, - 452007, India

(3) Space Research Centre, Polish Academy of Sciences, Bartycka 18A, 00-716 Warsaw, Poland

(4) Goddard Space Flight Centre, NASA, Washington, USA

## Abstract

Spotless days (SLDs) as well as CMEs in the decay phase of the solar cycle are believed to be a good predictor of the forthcoming cycle. A sequential increase in SLDs is observed since cycle 21, and cycle 24 has highest number of SLDs (since cycle 14), regarding it as one of the weakest cycles, which offer a unique opportunity to probe the CME characteristics that occurred during SLDs (hereafter $CME_{SLD}$), in a statistical sense. Here, we investigate the $CME_{SLD}$ during the descending phases of solar cycles 23 (2004-2008) and 24 (2015-2019). The fraction of CMEs that occurred on SLDs is found to be 14 and 11%, for cycles 23 and 24, respectively, compared to the total CMEs that occurred in the aforementioned durations. An increase in the number of $CME_{SLD}$ is revealed after 2004. Further, $CME_{SLD}$ that occurred on the visible side of the solar disk are found to be slower, smaller in width, and carrying low Kinetic energy and mass compared to the entire population of CMEs. The distribution of annual evolution of speed, angular width and acceleration of the $CME_{SLD}$ with the CMEs that occurred on the non-SLD days (hereafter $CME_{Non\text{-}SLD}$) for the descending phases shows that the $CME_{SLD}$ are different from the $CME_{non\text{-}SLD}$ in terms of characteristics (such as speed, width and acceleration) which also exhibits solar cycle dependence. A comparative analysis of $CME_{SLD}$ kinematics of cycle 23 and 24 shows that the weakest cycle 24 has wider and more massive events. In contrast, other parameters of $CME_{SLD}$ such as speed, acceleration and Kinetic energy do not have a disparate nature. $CME_{SLD}$ in both the cycles 23 and 24 are similar in nature in a statistical sense. Therefore, this investigation suggests that SLDs, and hence the sunspot number, may not be a sufficient candidate to predict the solar eruptive activities (e.g. CMEs). On the other hand, our analysis of the relation between the strength of the geomagnetic storm and probable candidate parameters revealed Dst index to be very well correlated with the product of V and Bz.

# 1. Introduction

The study of solar irradiance and eruptive activity is extremely important in context to their relationship with solar dynamo and **space weather changes**. (Haigh 2007, Charbonneau 2010). The eruptive activities on the sun, including solar flares and CMEs, influence space weather (Haigh 2007, Charbonneau 2010, Jain1986, Pulkkinen 2007, Pesnell 2012, Tsuda 2015, Kleimenova et al. 2018, Baker et. al. 2018, Labonville et al. 2019). It is well accepted that magnetic flux and helicity is released during the solar eruptive activities, such as CMEs and flares (Rust et al. 1994, Low 1996, Bao et al. 1998, DeVore 2000, Cliver et al. 2002, Liu et al. 2006, Démoulin 2007, Zhang et al. 2008). Active regions are one of the primal source locations of CMEs and flares, however, frequency and magnitude depends on their magnetic configuration, shear, flux and available free magnetic energy (Choithani et al. 2018, Parker 1979, Kulsrud 1998, Biskamp et al. 2000, Priest et al. 2000, Priest et al. 2002, Compagnino et al. 2017, Vemareddy 2021).

Notwithstanding the fact that CME occurrence rate and sunspot numbers are strongly correlated, from 2004 (SOHO/LASCO data) a significant enhancement in CME counts is observed in comparison to that seen in the sunspot number (Gopalswamy 2006, Petrie 2015, Michalek et al. 2019, Mishra et al. 2019). Chen et.al.2011 studied the location of 224 CME for year 1997-1998 and concluded that about 63% of the CMEs are related with active regions, at least about 53% of the active regions produced one or more CMEs, and particularly about 14% of ARs are CME-rich (3 or more CMEs were generated during one transit across the visible disk). Michalek et al. (2019) concluded that the decrease in heliospheric pressure (magnetic + plasma) and change in magnetic pattern of solar corona may be the reason to enhance the CME's rate occurrence. According to the Petrie (2015) and Mishra et al (2019), this increment in CME count is because of the large numbers of smaller ejections from higher latitude. On the other hand, Gopalswamy et al. (2006, 2003a, 2003b), also concluded that this enhancement is due to the CMEs which cannot be directly associated with sunspots. Further Mishra et al (2019) also confirm that the active sunspot regions are a subset source location of the CMEs.

Eruptive activities which originate outside the sunspot region are scarce. Dodson and Hedeman (1970) identified only 7 % of all flares during the period of 1956-1968, which are emitted from the plage regions with small or not identifiable spots. Altaş (1994) in the period 1947- 1990 observed only 2% of such flare events. Further, CME event that are not emitted from the sunspot vary in the range of 15% (during 1996-1998 - Subramanian et al. 2001), 21% (during 1997 to 2001 -Zhou et al. 2003), or 37% (1997 – 1998 -Chen et al. 2011 and Wang et al. 2011). Wagner (1984) used Skylab (1971-1974) and SMM (1979-1981) observations to identify such events (not associated with any feature such as flare, prominence etc.) and they found 50 and 67% respectively. For Solar cycle 23 (1996–2001), Vršnak et al. (2005) observed 545 pairs of CME-flare which has its source as disappearing filaments. Gopalswamy (2006); for the slightly extended period (1996 to 2006) concluded that the High Latitude CMEs are associated with (polar crown filaments) spotless eruptions. Ma et. al. (2019) claimed such events to be around 33% of the total CMEs that occurred in the first 8 months of year 2009 and termed them as stealth CMEs. They further found that, these stealth CMEs are relatively slow CMEs (v< 300 km $s^{-1}$) and exhibit rather gradual acceleration. D'Huys et al. (2014) studied 40 stealth CMEs during the year 2012.

However, several of such investigations correspond to the $CME_{SLD}$ that occurred in the rise phase up to the maximum of the solar cycle. These events contribute upto 37% of the total population of the CME. On the other hand, St. Cry and **Webb** (1991) noted more than 50% of such events during the descending phase solar cycle 21 (1984-1986). It is well known that solar emissions from non-active region locations are triggered by several phenomenon's such as activation and disappearance of filaments, newly appearing flux region, shear, propagation of slow mode of waves from another active region, large arcade and its instability etc. (Dodson and Hedeman 1970, Altaş 1994, Moore et al.1980, Rausaria et al.1992, Šeršeň et al.1993, Jain et al. 2011, Awasthi et al. 2014). Generally, the complex density distribution of the magnetic (current) helicity triggers the solar emission in the corona leaving negligible observational signatures in the photosphere. It has been shown that the released magnetic flux and helicity in the descending phase provides seed for the next solar cycle (Low 1996, Lamy et al. 2019, Kakad et al. 2019). Recently, Burud et al., (2021) showed that the number of spotless days (SLDs) observed during the descending phase of a given cycle is a potential candidate to forecasting the amplitude of the next solar cycle. Further, CMEs in the descending phase of the solar cycle are also believed to contain information of the forthcoming cycle (Hawkes et al. 2018). Therefore, in this paper, we investigate detailed properties of CMEs occurred on SLDs during descending phase of solar cycle 23 and 24 with the aim to identify the characteristics differences with $CME_{non\text{-}sld}$ and further improve their significance as proxy of the next solar cycle.

Table 1: Previous results of the study of Solar emission that occurred from the non-active regions

| Period of Study | Type of emission (no. of events) | % of the event over total population | Author | Methodology used |
|---|---|---|---|---|
| 1947- 1990 | Flare | 2 | Altaş (1994) | |
| 1956-1968 | Flare | 7 | Dodson and Hedeman (1970) | |
| 1971-1974, 1979-1981 | CME (58 and 87) | 50- 67% | Wagner (1984) | Solar maximum mission spaceraft and HAO Mark III K coronameter |
| 1984-1986 | CME (73) | < 50 | St. 3Cry and webb (1991) | Solar maximum mission spaceraft and HAO Mark III K coronameter |
| 1996-1998 | CME (32) | 15% | Subramanian, and Dere (2001) | on-disk signatures in images from the EUV Imaging Telescope (EIT) |
| 1997- 1998 | CME | 37% | Wang et al., (2011), Chen et al. 2011 | magnetic synoptic charts of Michelson Doppler Imager (MDI) |
| 1997 -2001 | CME (197) | 21% | Zhou et al. (2003) | EIT data |
| 1996–2001 | CMEs-flare (544) | | B. Vršnak, D. Sudar, and D. Ruždjak (2005) | Goes and SOHO |
| 2009 (January-August) | CME | 33% | Ma et. al. (2010) | EUVimager(EUVI)andthecoronagraphsCOR1 and COR2 on board STEREO. |
| 2012 | CME (40) | 2.5% | D'Huys et al. (2014) | |

We use the SOHO/LASCO data set to study and compare the characteristics of CME over the descending phase of the solar cycle 23 (2004-2008) and 24 (2015-2019). To the best of our knowledge, this is the very first investigation that utilizes such a large data set for obtaining the CME properties on SLDs and their role in inferring the properties of the next cycle. This paper is arranged as follows. In section 2, we describe data sources and in section 3 we present results. A brief summary along with discussion has been presented in section 4.

## 2. Sources of Data:

We identify SLDs using the sunspot number records made available at the Solar Geophysical Data (SGD) reports ftp://ftp.ngdc.noaa.gov/STP/swpc_products/weekly_reports/PRFs_of_SGD/ during descending phase of solar cycle 23 (2004 to 2008) and 24 (2015-2019). The flare location and its GOES intensity class have been obtained from the SGD reports, and cross verified with the Hinode flare catalogue (https://hinode.isee.nagoya-u.ac.jp/flare_catalogue/). The CME data has been obtained from the SOlar Heliospheric Observatory with Large Angle and Spectrometric Coronagraph Experiment (SOHO/LASCO) CME catalogue (URL: https://cdaw.gsfc.nasa.gov/CME_list/). We use the CME properties such as linear speed, angular width, acceleration, mass and kinetic energy reported in this catalogue in the current study. Further, we classify the reported CMEs as front side or back side by analyzing Extreme Ultra-Violet (EUV) images taken in 195Å by SOHO or Solar Dynamics Observatory (SDO) satellites. If the source location of the CME is found on the visible side of Sun, it is considered as front side CME. It may be noted that the 195Å image data set from EIT/SOHO (Delaboudiniere et al. 1995) is not available from January 2017. Therefore, we employ SDO's Atmospheric Imaging Assembly (SDO/AIA**, Lemen et al. 2012**) observations taken in 193 Å for the period January 2017 to December 2019. The temporal cadence of EIT/ SOHO and AIA/ SDO is 10-20 min and 12 seconds, respectively. The prominence and filament data **have** been taken from Heliophysics Feature Catalogue (HFC), at URL: http://voparis-helio.obspm.fr/hfc-gui/index.php in association with the CAII K3 Spectroheliograph of the Meudon Observatory. To identify the relation between solar wind plasma parameters and Dst indices, the list of hourly values of solar solar wind parameter provided by National Geophysical Data Center ( https://omniweb.gsfc.nasa.gov/form/dx1.html) and Dst indices given by the World Data Center for Geomagnetism, Kyoto, Japan through its world wide web (https://wdc.kugi.kyoto-u.ac.jp/index.html) were used.

## 3. Analysis and Results:

### 3.1 Spotless Days (SLDs)

We found that SLDs show an increasing trend as we approach to the solar minimum. Therefore, number of SLDs increase during descending phase of a given solar cycle as shown in Fig. 1. The Fig. 1 shows a continuous progression in SLD number during descending phase of cycle 21 to 24. It may be noted that SLDs has been increasing from cycle 21 through 24. The total number of SLD in cycle 21,22,23 and 24 are found to be 254,289, 537 and 931, respectively. Therefore, the relative increase in total number of SLDs with respect to previous cycle is estimated to be 1.13 ( $\frac{SLD_{22}}{SLD_{21}} = \frac{289}{254}$), 1.98 ($\frac{SLD_{23}}{SLD_{22}} = \frac{573}{289}$)and 1.59( $\frac{SLD_{24}}{SLD_{23}} = \frac{913}{573}$) . The cycle 24 shows the highest numbers of SLD compared to the previous three cycles (see Fig. 1). A total of 514 SLDs have been observed during the descending phase of solar cycle 23, while 623 SLDs (21% larger than former) in the descending phase of cycle 24. Continual increase in the SLDs count for the descending phase of cycle 21 to 24 is found to be associated with a sequential weakening of magnetic field generation in the solar interior (Georgoulis et al. 2019, Kakad et al. 2019). The longest recorded period of consecutive spotless days was 52 (July 21 – September 10, 2008) in cycle 23 and 36 days (May 19 – June 23, 2019) in cycle 24.

From fig. 1, SLDs recorded during the minimum year of both the cycles (i.e. 2008 and 2019) are found to be nearly equal (~273). Although, the monthly mean sunspot numbers 3.52 and 2.8 for the minimum year 2008 and 2019 respectively, slightly higher for year 2019. Woods et al. (2022) noted similarities for the minimum of cycle 23 and 24 for the data (SORCE mission) of total solar irradiance and solar spectral irradiance.

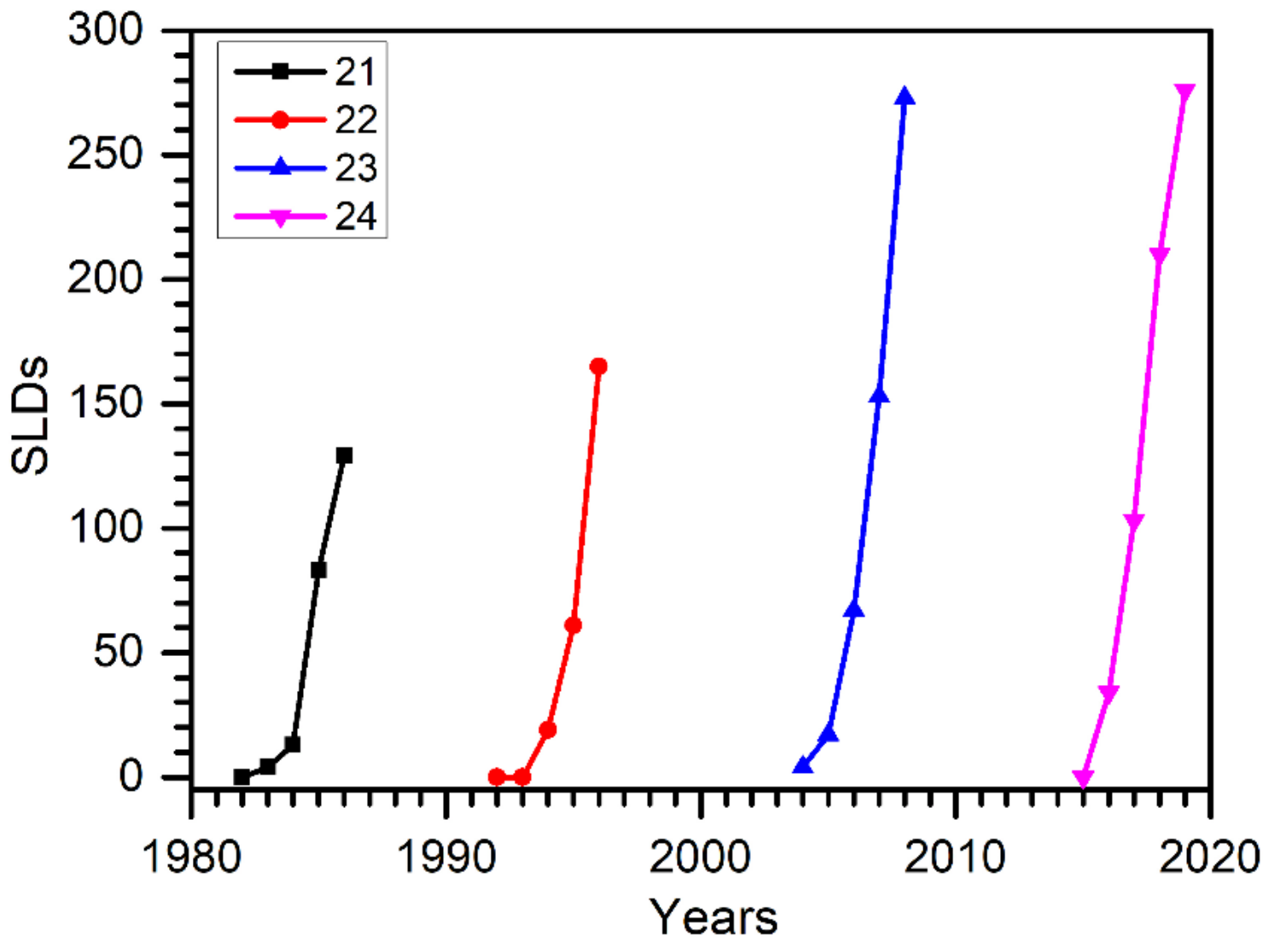

Figure 1. The year wise variation of SLDs during the descending phases of cycle 21-24.

### 3.2 Characteristics of CMEs which occurred on SLDs ($CME_{SLD}$ ):

#### 3.2.1 Front side CME events on SLDs:

For $CME_{SLD}$, we investigated the time period covering years 2004-2008 and 2015-2019, respectively, defined as the descending phase of solar cycles 23 and 24 in Burud et al. (2021). According to the CDAW CME catalogue, we identified a total of 5702 CMEs (1343 CMEs on SLD ) during cycle 23 and 5262 CMEs (933 CMEs on SLD) during cycle 24, corresponding to an - 8.3% decrease (Table 1).

Next, these events are classified as a backside and frontside events by visually identifying the source location of the CMEs that are observed in the C2 field-of-view of LASCO coronagraph, on the solar disk by utilizing the co-temporal EUV images obtained by EIT/SOHO. In particular, the dimming and position angle in C2 image have been correlated with EIT image for the correct source identification. Fig. 2 exemplifies the cases of frontside (upper panel) and backside (lower panel) CME events. The left image in this panel is the EUV image obtained from EIT telescope; the middle image shows the running difference LASCO/ C2 observation while the right image shows the combination of C2 and EIT images. Arrows on the images denote the identified source locations of CMEs. In this way, with the help of EUV (EIT-195Å and AIA/SDO-193Å) images and movies, we have identified a total of 800 and 589 front side $CME_{SLD}$ events during the descending phase of cycle 23 and 24, respectively. Frontside $CME_{SLD}$ events amounted to 14 and 11% of total CMEs observed during the descending phase of cycle 23 and 24, respectively.

Table 2: Summary of CME events occurred during the descending phase of solar cycle 23 and 24 as per CDAW catalog

| Cycle-23 | | | | | Cycle-24 | | | | |
|---|---|---|---|---|---|---|---|---|---|
| Year | SLD | CME (All) | CME on SLD (All) | Front-side $CME_{SLD}$ (fraction in %) | Year | SLD | CME (All) | CME on SLD (All) | Front-side $CME_{SLD}$ (fraction in %) |
| 2004 | 04 | 1102 | 14 | 04 (0.36) | 2015 | 00 | 2058 | 00 | 00 |
| 2005 | 17 | 1249 | 61 | 39(3.12) | 2016 | 34 | 1393 | 96 | 61(4.37) |
| 2006 | 67 | 1046 | 164 | 100 (9.56) | 2017 | 103 | 786 | 219 | 155 (19.72) |
| 2007 | 153 | 1442 | 493 | 336 (23.3) | 2018 | 210 | 476 | 224 | 119 (25) |
| 2008 | 273 | 863 | 611 | 321 (37.19) | 2019 | 276 | 549 | 394 | 254(46.26) |
| Total | 514 | 5702 | 1343 | 800 (14) | Total | 623 | 5262 | 933 | 589 (11.1) |

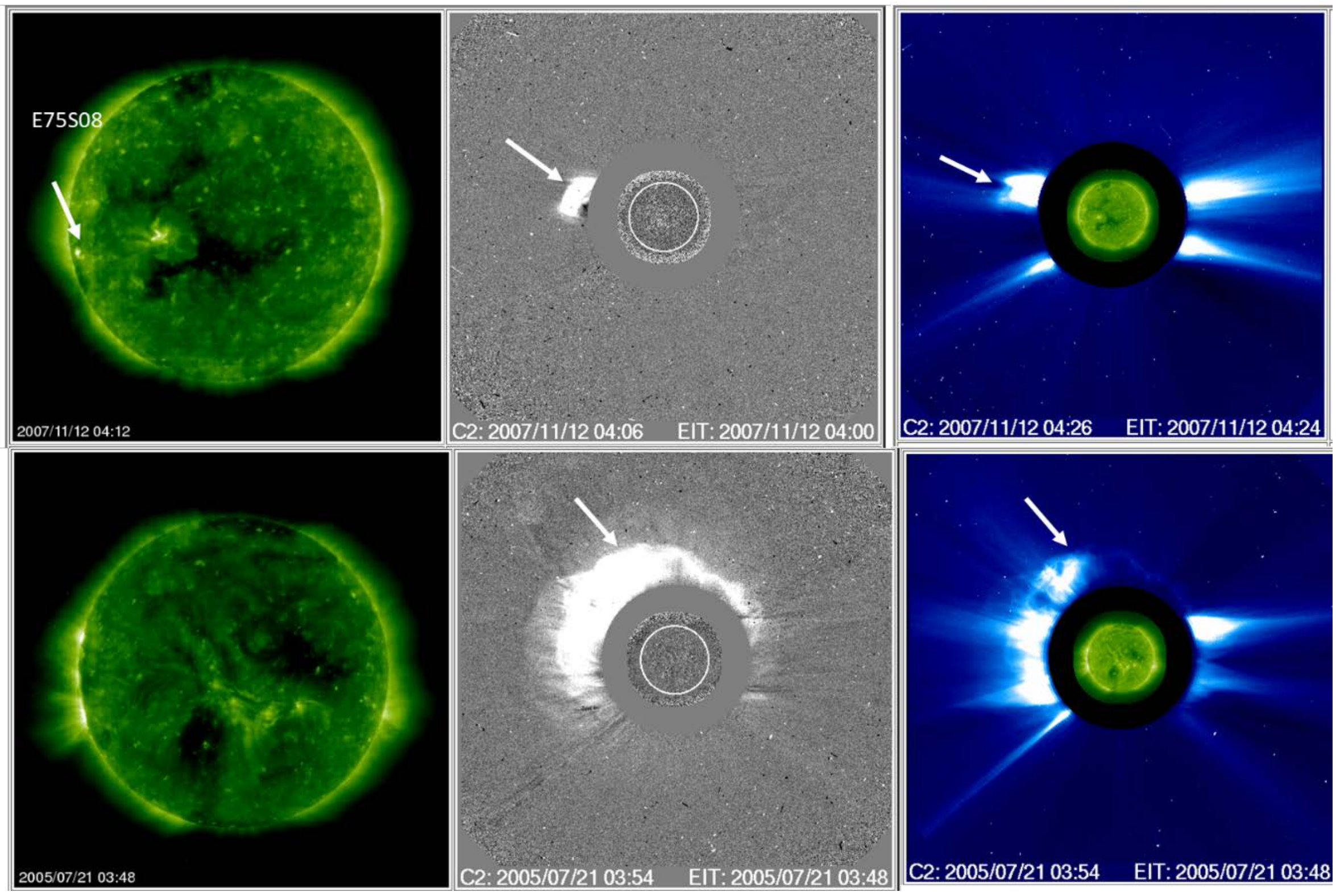

Figure 2: Upper panel shows front-side $CME_{SLD}$ event of 12 November 2007 with unambiguously identifiable source location on the visible disk of the Sun, while the lower panel shows $CME_{SLD}$ event of 21 July 2005, for which the source location could not been identified, and hence considered as back-side event.

### 3.2.2 $CME_{SLD}$ activity in cycle 23:

Table 2 presents that 800 $CME_{SLD}$ observed during 514 SLDs occurred in the descending phase of solar cycle 23 (occurrence rate $CME_{SLD}$/SLD of 1.55). The annual number of SLD, $CME_{SLD}$ and fraction of $CME_{SLD}$ has an increasing trend towards the solar minimum (see table 2). Fig. 3a illustrates a strong correlation (R=0.88) between the number of $CME_{SLD}$ and SLD count during each year of descending phase. The average values of $CME_{SLD}$ parameters viz. linear speed, angular width, acceleration, Kinetic energy and mass for the period of 2004-2008 were 270 km/s, $23.46^0$, 7.84 km/s$^2$, $3.8\times10^{29}$ erg and 6.5 $\times10^{14}$ gm, respectively. Mittal & Narain (2009) found that the average values of linear speed, angular width, acceleration, Kinetic energy and mass for all CME observed for a period (1996-2007) were 435 km/s; $41^0$; $1.9 \times 10^{30}$ erg and $1.2 \times 10^{15}$ gm respectively. This implies that $CME_{SLD}$ are slower, smaller in width, carries less Kinetic energy and mass **than CMEs**. The range of the evolution of the average annual speed (250 to 360 km/s) and angular width ($19^0$ to $45^0$) of $CME_{SLD}$ event is always less than that for the all CME (speed = 300 to 450 km/s ; angular width=$30^0$ to $60^0$) parameters over the same period. In contrast, CMEs occurred in the beginning cycle 24 (in year 2009) exhibited slightly higher angular width (40°) and speed (300 km/s) (Ma et al. 2010).

Table 3 presents the annual average distribution of these $CME_{SLD}$ parameter. The annual average can be calculated by considering the total number of $CME_{SLD}$ in corresponding year; then averaging the parameter with respect to this number for the same time period. Table 3 illustrates the annual average linear speed (c.f. Fig. 3b) and angular width of $CME_{SLD}$, which gradually decreased (from 360 to 250 km/s, and 44 to 20 degrees, respectively). Further we have delineated the $CME_{SLD}$ in three groups with respect to linear speed: slow (v<300 km/s); intermediate ($300 \leqslant v \leqslant 600$ km/s) and fast (v>600 km/s). We found 539 (67 %), 241 (30%) and 19 (2.3%) events with slow, intermediate and fast speed respectively. This reveals that a significant fraction of the $CME_{SLD}$ are of slow in nature. These slow CMEs experience a drag **force** from their interaction with the ambient solar wind, which can cause their speed to either decrease or **remain** constant during propagation (Webb et al. 2012). **Out of the 539 slow $CME_{SLD}$, 69 % of the slow CMEs have accelerated, while the rest 31% decelerated**. This result is in good agreement with Yashiro et. al.,(2004) and Du et. al.,(2021). The 63% of $CME_{SLD}$ events having intermediate speed exhibited acceleration as they propagate while the rest 35% decelerated. On the other hand, an equal number (**9**) of fast $CME_{SLD}$ exhibited acceleration and deceleration. Kinetic energy (fig. 3c) and mass (fig. 3d) distribution shows the continuous increase up to 2006; although an abrupt decrease is observed during 2007, which subsequently increased again during the minimum year 2008. It is to note that the Kinetic energy and mass data is available only for 272 events in the CDAW catalogue.

In addition, as per the **scheme of** Gao, P.X., and Li (2009), $CME_{SLD}$ events are further subdivided into two categories based on the signs of accelerating (positive) and decelerating (negative) events.

The annual average acceleration as seen in fig. 3e and Table 3 shows the continuous increment during the descending phases of the solar cycle except the year 2005 (for acceleration) and 2005, 2007 (for deceleration). The digits above each of the bar in fig.5e represent the total number of events recorded in that year.The number of acclerated $CME_{SLD}$ is almost double than that of the decelerated events. The minimum year 2008 has the same average value (32.30 km/s$^2$) for both types (accelerating/decelerating) of events. Our result is in agreement with Compagnion et. al.(2017) , who reported average annual acceleration to be negative (positive) for the year which has higher (lower) activity. **Slow CMEs** (less than 400 km/s in outer corona ) are effected by the propelling force rather than the retarding force (a drag force due to the solar wind) and vice versa (Webb et al. 2012, Gao, P.X. and Li 2009). In our investigation, most of the $CME_{SLD}$ events have speeds less than 400 km/s and hence they are accelerated. Our results are in agreeement with the previous results (st Cyr *et al.*, 2000; Yashiro *et al.*, 2004; Gao, P.X. and Li 2009, Kay et.al.2022).

Table 3: The annual variation of parameters of the front-side CME$_{SLD}$ during the descending phase of solar cycle 23

| Year | Mean A.W. (degree) | Average Speed (km/s) | Average mass ×10$^{14}$(gm) | Average Kinetic energy ×10$^{29}$(erg) | Average acceleration (km/s$^2$) | Average dceceleration (km/s$^2$) |
|---|---|---|---|---|---|---|
| 2004 | 21.5 | 360.5 | 2 | 0.67 | 2.4 (01) | -8(02) |
| 2005 | 44.17 | 342.02 | 5.3 | 1.35 | 37.29(24) | -121(06) |
| 2006 | 36.17 | 290.16 | 9.52 | 6.05 | 18.66(60) | -33.80(36) |
| 2007 | 19.92 | 249.74 | 3.81 | 1.13 | 21.87(214) | -18.91(118) |
| 2008 | 20.71 | 275.3 | 7.71 | 5.69 | 32.34(219) | -32.26(99) |

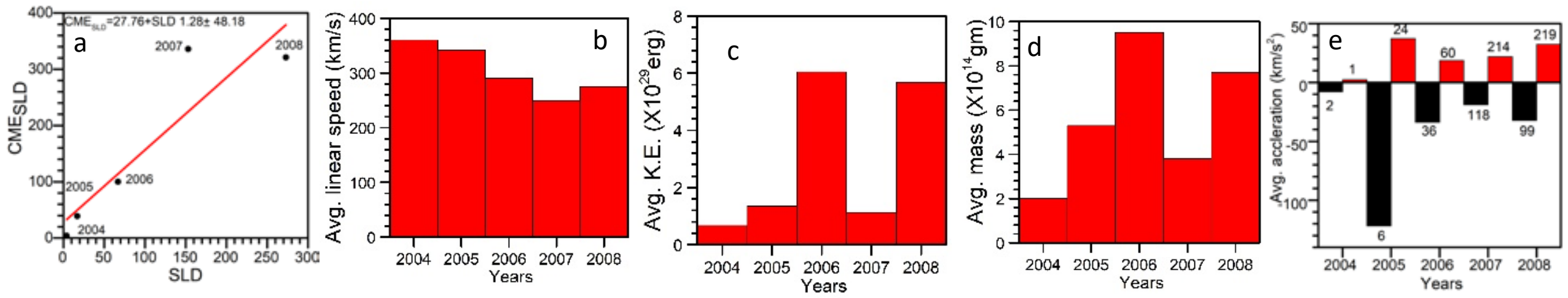

Figure 3 : The distribution of kinematic properties of the CME$_{SLD}$ in a given year of the descending phase of cycle: (a) distribution of SLDs and CME$_{SLD}$ (b) average linear speed (c) average Kinetic energy (d) average mass and (e) average acceleration ( red) and decelerration (black colour).

In order to delineate the parameters of CME$_{SLD}$ events from that of all CMEs that took place in the same period of analysis with taking into account the exclusion of selection bias, we adopted the same methodology used by D'Huys et al. (2014). In this regard, since in the year 2004, we observed only **4** CME$_{SLD}$ out of 1102 CMEs, we prepared 717 samples of **4** CMEs, selected in a random manner (i.e 65% of total CME events observed by the CDAW catlogue) and calculated mean speed for each of the set. Fig. 4 shows the plot of the distribution of mean speed, angular width and acceleration of these randomly chosen 717 samples. We find that the speed range of the randomly selected samples is in between 179-926 km/s, which is generally higher than the average speed of CME$_{SLD}$ events (red colour). The average speed of the CMEs in the sample (428 km/s; blue colour) amounts to be ~ 15% higher than that of the CME$_{SLD}$ (360.5 km/s; red colour; c.f. table 4). Similar inferences can be made from the analysis done for other years (c.f. table 4). Therefore, this exercise reveals that the characteristics of CME$_{SLD}$ are not strongly affected by the selection bias, and may have a different kinematical nature than the CME events that occurred on the non-SLD days. We also found the annual averaged width of CME$_{SLD}$ to be little lower to that obtained for the samples. **It is evident from figs. 4a and 4b and table 4 that** the CME$_{SLD}$ event possess lower speed and width except minimum year. During the descending phase from 2004 to 2008, the rate of decrease in CME samples, average speed (20 to -3%) and width (64 to -3%) is higher when compared with that of CME$_{SLD}$. **Sunspot number, a proxy of the Solar cycle plotted green in fig. 4.** Fig.4c reveals that the average acceleration has anticorrelation with the solar cycle progression. **While the $CME_{SLD}$ events exhibited both the acceleration and deceleration during propagation, majority of the events exhibited acceleration**. Hence our result strongly support the result of Compagnion et al. (2017).

Table 4: The characteristics of the distribution of the CMEs having the sample size same as $CME_{SLD}$ observed in that particular year.

| Year (no. of the $CME_{SLD}$) | No. of the set of the CME sample events | Speed (km/s) | | Width (degree) | | Acceleration (km/s2) | |
|---|---|---|---|---|---|---|---|
| | | Avg. | Range | Avg. | Range | Avg. | Range |
| 2004 (04) | 717 | 428 | 179-926 | 60.6 | 17-233 | | |
| 2005 (39) | 816 | 429.5 | 409-448 | 60.2 | 13-114 | 1.56 | -729 to 516 |
| 2006 (100) | 684 | 320 | 305-331 | 38 | 26-54.3 | 5.57 | -365 to 694.4 |
| 2007 (336) | 944 | 255 | 247-261 | 23.4 | 20-28 | 5 | -231.6 to 352 |
| 2008(321) | 571 | 267 | 257-277 | 20 | 17-24 | 7.99 | -331.2 to 389 |

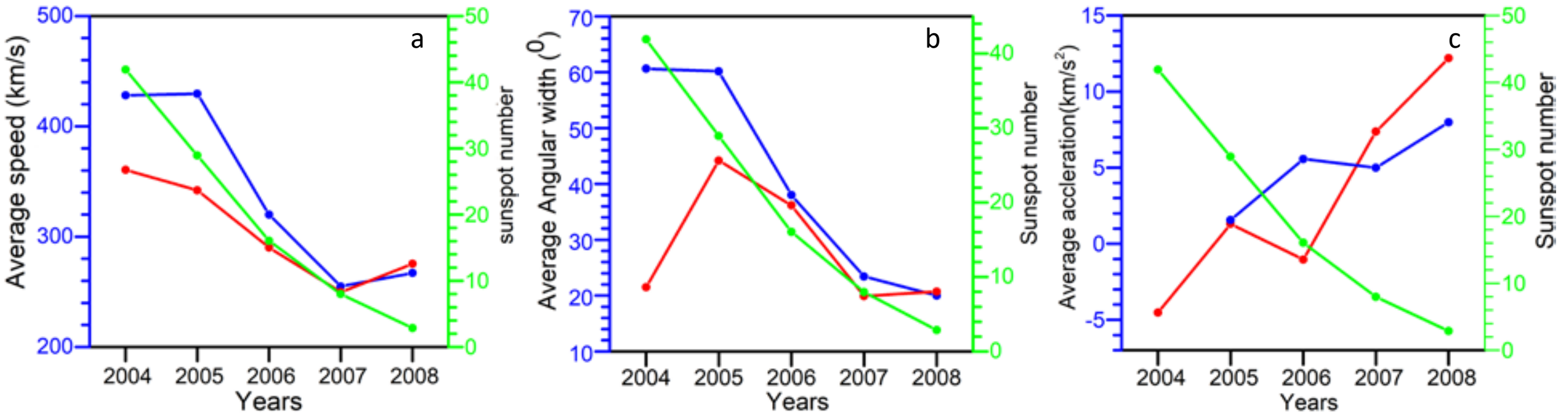

Figure 4. Comparison of average speed (a), Width (b) and Acceleration (c) distribution of the set of the randomly selected samples of the total annual number of CMEs events (blue colour) and $CME_{SLD}$ (red colour). Sunspot number, a representative of Solar cycle, is plotted in green colour.

### 3.2.3 $CME_{SLD}$ activity in cycle 24:

Table 2 **presents** 589 CMEs **associated with** 556 SLD**s observed** during the descending phase of solar cycle 24 (2015-2019). Similar to cycle 23, number of SLD, $CME_{SLD}$ and fraction of CMEs were increasing towards sunspot minimum (see table 2). The SLD and $CME_{SLD}$ has a high correlation with occurrence rate of 1.05 (*c.f.* fig. 5a). The average values for the $CME_{SLD}$ parameters (over the period 2016-2019; as the year 2015 has zero SLD) linear speed, angular width, acceleration, Kinetic energy and mass were 266 km/s, $31.90^0$,11.67 km/$s^2$, $2.27\times10^{30}$ erg and 7.22 $\times10^{14}$ gm respectively. Table 5, represents the annual average values of $CME_{SLD}$ parameters for the period 2015-2019. The average annual speed, width of the $CME_{SLD}$ ranging from 233 to 314 km/s, $22^0$ to $46^0$; while for all the CMEs (over the same period) it ranges between 250 to 330 km/s ; $38^0$ to $52^0$ respectively (Lamy et al. 2019). Hence it has been clear from Table 5 that , the $CME_{SLD}$ events has lower speed and angular width. The average annual linear speed (*c. f.* fig. 5b) and angular width (see table 5) of $CME_{SLD}$, is continuously decreases from 2016 to 2019. Among 589 $CME_{SLD}$, 406 ,162 and 20 **events** were categorized as the slow, intermediate and high speed respectively. Out of these 406 slow speed events; 272 (67 %) and 134 (33%) were accelerated and decelerated respectively. For event with intermediate speed; 105 (64%) and 57 (36%) were accelerated and decelerated respectively. On the other hand, an equal number of fast CMEs are accelerated and decelerated. We found Kinetic energy and mass values for only 300 events in the database. The **Figs.** 5c and 5d **show** the variation of average Kinetic energy and mass of the $CME_{SLD}$. Like cycle 23, Kinetic energy and mass shows the continuous increasing trend up to the year 2017 has a sudden drop for the year 2018 and again an increase for year 2019. The annual average acceleration continuously increases towards the solar minimum (see table 5). Similar to cycle 23, $CME_{SLD}$ events are divided as decelerating (black) and accelerating (red). A continous increasing trend has been observed in both types of acceleration up to the solar minimum. The digits above each of the bar in fig.5e represents frequency of events observed in that year. The accelererating events are almost twice as frequency as the decelerating event in the same year. While the annual average acceleration has positive values. The CME driving force is inversely proportional to the sunspot activity. Therefore, acceleration (deceleration) of CME event depends on the weak (strong) solar activity (Du 2021). While the mean values of CME speeds during SLD and non-SLD exhibited a declining trend in the descending phase of the solar cycle, a continual increase in the acceleration is spotted in general. This may possibly be associated with the effect of drag force exerted by the solar wind on slow CMEs (Kay et.al.2022)

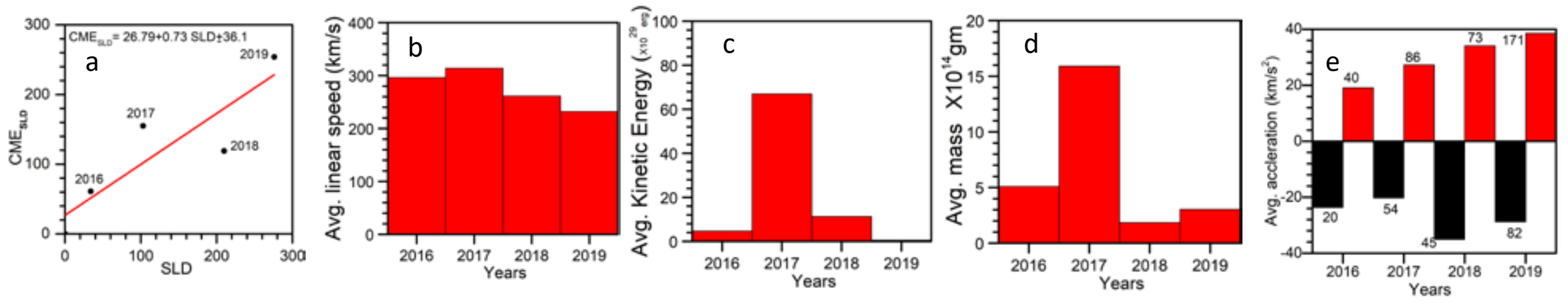

Figure 5: The distribution of kinematic properties of the $CME_{SLD}$ in a given year of the descending phase of cycle: (a) distribution of SLDs and $CME_{SLD}$ (b) average linear speed (c) average Kinetic energy (d) average mass and (e) average acceleration ( red) and decelerration (black colour).

Table 5: The annual variation of CDAW detected CME parameters on SLDs during the descending phase of solar cycle 24 (2015- 2019).

| Year | Average A.W. (degree) | Average Speed (km/s) | Average mass ×10^14(gm) | Average Kinetic energy ×10^29(erg) | Average acceleration (km/s²) | Average dcceleration (km/s²) |
|---|---|---|---|---|---|---|
| 2016 | 46 | 296.4 | 5.09 | 4.68 | 19.07(40) | -23.64(20) |
| 2017 | 44.19 | 313.79 | 15.9 | 67 | 27.28(86) | -20.25(54) |
| 2018 | 28.12 | 261.32 | 1.85 | 11.3 | 34.08(73) | -35.05(45) |
| 2019 | 22.79 | 232.16 | 3.04 | 0.56 | 38.55(171) | -28.73(82) |

Like the cycle 23, using the same methodology the comparison between the different characteristics of $CME_{SLD}$ and CME for the same year is done and the results are summarised in the table 6. Fig. 6 and table 6, demonstrate that the average annual value of speed, width of $CME_{SLD}$ follows the solar cycle; Sunspot number, a representative of Solar cycle, is plotted in green. While the acceleration has an anticorrelation with the solar cycle (see the fig. 6c). For the year 2019, the clear difference between speed, angular width and acceleration of two data sets (i.e. CMEs-blue and $CME_{SLD}$-red ) has been observed.

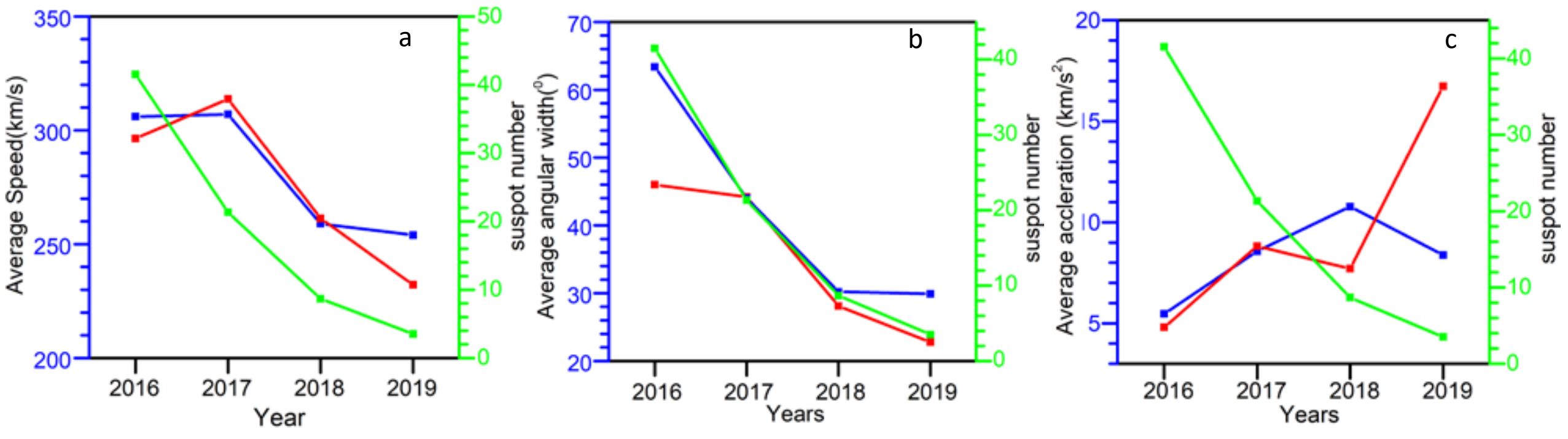

Figure 6.Comparison of average speed (a), Width (b) and Acceleration (c) distribution of the set of the randomly selected samples of the total annual number of CMEs events (blue colour) and $CME_{SLD}$ (red colour). Sunspot number, a representative of Solar cycle, is plotted in green colour.

Table 6: The characteristics of the speed distribution of the CMEs detected by the CDAW having sample size same as $CME_{SLD}$ observed in that particular year.

| Year (no. of the $CME_{SLD}$) | No. of the set of the CME | Speed (km/s) | | Width (degree) | | Acceleration (km/s²) | |
|---|---|---|---|---|---|---|---|
| | | Avg. | Range | Avg. | Range | Avg. | Range |
| | sample events | | | | | | |
| 2016 (96) | 905 | 306 | 265-367 | 63.36 | 42-87 | 5.47 | -705 to 425 |
| 2017(219) | 511 | 307 | 276-355 | 44 | 36-53 | 8.57 | -237.7 to 509.7 |
| 2018 (224) | 309 | 259 | 245-272 | 30.2 | 26-35 | 10.78 | -719 to 689 |
| 2019 (394) | 357 | 254 | 243-262 | 29.9 | 26-32 | 8.38 | -564.3 to 437.3 |

### 3.2.4 Geomagnetic storms associated with the $CME_{SLD}$:

The interplanetary manifestations of CMEs, i.e. magnetic cloud and Corotating Interaction Region are the most geoeffective interplanetary phenomena (Gonzalez, et al 1994, Yermolaev& Yermolaev 2010). On the **other hand**, southward component of **the interplanetary magnetic field** with magnitude >10 nT with a long-duration (few or more hours) impact,will affect Earth's geomagnetic field. This decrease in the Earth's geomagnetic field is recoreded as the disturbance storm time index (Dst). Further, the solar wind parameters such as interplanetary magnetic field , plasma temperature, plasma density, plasma speed, plasma pressure, etc., are statistically correlated with the Dst index (Maltsev, Y.P. and Rezhenov, B.V., 2003).

The **linear** speed **and** width of CME (if Earth directed) is a key factor to calculate the strength and time of occurrence of geomagnetic storm (Kim et al. 2017; **Dumbović** et al. al. 2015; Schwenn 2005; Zhang et al. 2007). The observations of solar wind plasma parameters at 1 AU, shows that even though the slow speed and narrow events can also reach to Earth, may produce geomagnetic storms (Kilpua et al. 2014; Nitta 2017).

Therefore, we have classified our data set of $CME_{SLD}$ in three different groups with respect to linear speed: slow ($v \leqslant 300$ km/s); intermediate ($300 < v \leqslant 600$ km/s) and fast ($v > 600$ km/s). **Both cycles** have upto 68 and 64% of slow and Intermediate speed $CME_{SLD}$ events has been accelerated during propagation respectively.

One of the probable reasons for the weak geomagnetic activity is the interaction of slow CMEs with the large-scale solar wind structures or preceding CMEs during the propagation. On the other hand, the induced electric field Ey = VBz and its influence on magnetospheric–ionospheric system also plays an important role in producing storms and substorms (Lakhina, G.S. and Tsurutani, B.T., 2016, Nikolaeva, N.S., Yermolaev, Y.I. and Lodkina, I.G., 2011). Despite, there is lack of one-to-one association of storm intensity with the flare and CME occurance (Dumbović et al, 2015, Liu et al., 2016). As listed in Table 7, a total of 13 (SC23) and 10 (SC24) events are probably responsible for producing the minor geomagnetic storms ($Dst_{min}$ between −30 and −86 nT).

Table 7 lists the corresponding $CME_{SLD}$ and IMF parameters, which seems to be responsible for the geoeffectiveness. The Bz value ranges from -4 to -11nT, solar wind speed from 350 to 676 km/s, while the **linear** speed of the $CME_{SLD}$ lies between 180 to 430 km/s. However, we did not find any significant statistical relation between CME parameters such as angular width, **linear** speed and speed at 20 solar radii. The **linear** speed of the $CME_{SLD}$ is less than or equal to the solar wind speed (400 km/s) **may** have been increased as they reach 1 AU (see table 7). The **linear** speed of CME depends on the free energy of the source and is modified by the interaction with the solar wind. The product V and Bz exhibited a good correlation with the Dst index as seen in the fig. 7 with a relation of

$$Dst = -0.01V\,Bz - 12.$$

Solar wind speed and the southward magnetic field component of CME determine the strength of the storm as given by equation 1, Gopalswamy et al. (2009).

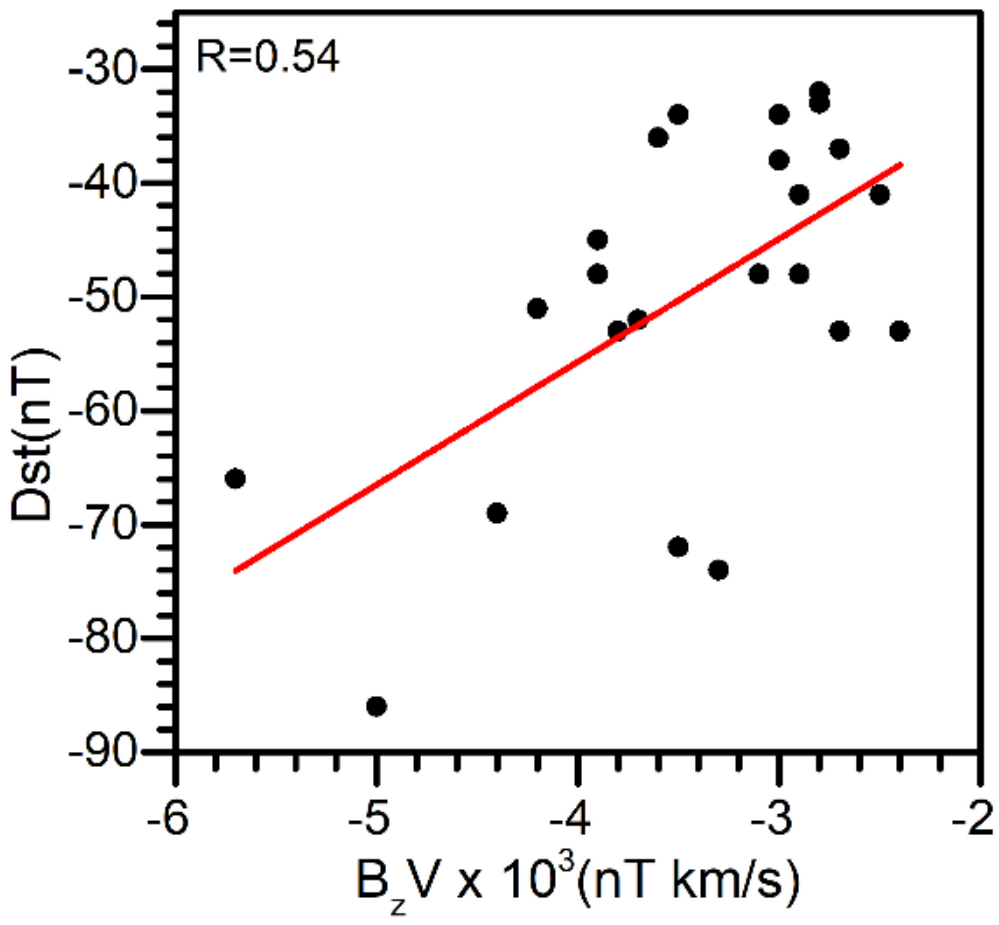

Figure 7 .The relation between the product of $B_Z$ (southward component of Interplanetary magnetic field) , Speed of the solar wind with the Dst index. They has good correlation coefficient (R=0.54).

Table 7 : The list of the Geomagnetic storms linked with $CME_{SLD}$ events and its parameters.

| $CME_{SLD}$ | | | | | IMF Parameters | | | |
|---|---|---|---|---|---|---|---|---|
| Date | Time (UT) | Angular width($^0$) | **Linear** speed (km/s) | Speed at $R_{20}$ (km/s) | Date | $B_z$(nT) | V(km/s) | $D_{st}$(nT) |
| 24-10-2005 | 02:54:05 | 34 | 159 | 566 | 24-10-2005 | -7.7 | 356 | -37 |
| 27-10-2005 | 18:54:05 | 29 | 188 | 844 | 27-10-2005 | -9.2 | 368 | -74 |
| 18-02-2006 | 11:54:04 | 37 | 242 | 0 | 18-02-2006 | -6.7 | 548 | -36 |
| 17-05-2006 | 08:30:05 | 7 | 329 | --- | 17-05-2006 | -8.1 | 376 | -38 |
| 06-08-2006 | 15:30:04 | 29 | 252 | 0 | 06-08-2006 | -9.2 | 429 | -48 |
| 12-02-2007 | 09:30:04 | 39 | 402 | 350 | 12-02-2007 | -4.8 | 606 | -41 |
| 20-03-2007 | 14:26:04 | 23 | 349 | 188 | 20-03-2007 | -9.6 | 373 | -72 |
| 23-10-2007 | 14:30:04 | 8 | 429 | 0 | 23-10-2007 | -5.6 | 429 | -53 |
| 11-01-2008 | 00:30:04 | 10 | 295 | 440 | 11-01-2008 | -6.9 | 443 | -34 |
| 07-03-2008 | 02:54:05 | 33 | 205 | 181 | 07-03-2008 | -10.6 | 481 | -86 |
| 11-07-2008 | 09:54:17 | 36 | 186 | 307 | 11-07-2008 | -7.2 | 498 | -34 |
| 07-08-2008 | 00:30:04 | 15 | 225 | 1131 | 07-08-2008 | -4.3 | 653 | -33 |
| 02-09-2008 | 08:06:04 | 12 | 306 | 940 | 02-09-2008 | -8.7 | 488 | -51 |
| 19-07-2017 | 14:00:05 | 8 | 194 | 708 | 19-07-2017 | -5.6 | 517 | -36 |
| 09-10-2017 | 05:36:05 | 104 | 327 | 416 | 09-10-2017 | -7.2 | 383 | -53 |
| 03-11-2017 | 23:12:12 | 29 | 208 | 384 | 03-11-2017 | -11 | 405 | -82 |
| 03-12-2017 | 02:12:05 | 11 | 268 | 420 | 03-12-2017 | -11.1 | 354 | -51 |
| 18-04-2018 | 07:24:05 | 9 | 323 | 229 | 18-04-2018 | -13.5 | 426 | -78 |
| 19-09-2018 | 08:24:05 | 41 | 406 | 1198 | 19-09-2018 | -7.1 | 365 | -53 |
| 02-11-2018 | 16:00:06 | 46 | 276 | 0 | 02-11-2018 | -7.1 | 412 | -57 |
| 04-08-2019 | 17:36:05 | 37 | 431 | 136 | 04-08-2019 | -7.9 | 489 | -53 |
| 28-08-2019 | 09:48:05 | 84 | 167 | 187 | 28-08-2019 | -6 | 632 | -52 |
| 26-09-2019 | 01:36:14 | 26 | 125 | 352 | 26-09-2019 | -5.7 | 554 | -56 |

### 3.2.5 Association of $CME_{sld}$ with other solar activity:

CME events are generally associated with other solar activity such as flare, eruptive prominences, filament disappearances, helmet streamer disruptions and blowouts (Munro et al. 1979; Kahler 1992; Hundhausen 1993; Harrison 1991). We check different datasets described as in section 2 for the association of solar activity with corresponding $CME_{SLD}$. We have tried to establish a correlation only if both the activities occur within a time difference of 1 hour. Further we studied characteristics of such events associated with $CME_{SLD}$ events.

It is believed that the flares, which are not accompanied by CME events, have higher temperature, since the amount of magnetic energy release is used to heat the plasma predominately. On the other hand, for the CME associated flares events, large part of energy released is utilized in accelerating the ejecting particles (Yashiro et.al. 2006). Vršnak, Sudar, and Ruždjak (2005) and Bein et al. (2012) observed the similarities in the acceleration of the CME event emitted from spotless regions and active-region CMEs associated with weak flares of SXR-class A, B and C and concluded weakest acceleration. With the reference to this we studied the flare activity that is associated with the $CME_{SLD}$. A total 94 and 226 flares (⩽C class) occurred on SLD during the descending phase of cycle 23 and 24, respectively. Out of this, the source location of only 26 (56) flares have been identified on the visible side of the solar disk for cycle 23 (24). Further, all these flares have not been found to be accompanied by CMEs and vice versa. As listed in table 7, only **6** and **9** B-class flares were associated with the $CME_{SLD}$ in cycle 23 and 24 respectively. Usually, the CMEs associated with the flare with high flux (⩾ C class) have larger mass, speed, and width while low flux flare (⩽C class) do not show any clear trend (Lamy et al. 2019). Similarly, our data set of Flares on SLDs, (Table 8) do not shows any systematic trend in its parameters for both cycles.

From the set of 800 (589) $CME_{SLD}$ events, only 47(22) was associated with prominences in cycle 23 (24). Munro et al. (1979) and Gopalswamy et al. (2003b), have shown that the prominence with minimum height of 1.2 solar radii can produces CME. During the rising phase (1996 to 1998) of solar cycle 23, Subramanian & Dere (2001) found 15% of such CMEs events were linked with prominence eruptions. As shown in table 9, $CME_{SLD}$ events associated with prominence has higher average angular width and linear speed (c.f. table 3, 5 and 9). Current results are in agreement with the pervious result of Lamy et.al. (2019) who studied the properties of CME event associated with prominence eruptions for cycle 23 and 24. The numbers of $CME_{SLD}$ associated with the prominence has anticorrelation with the solar cycle; while angular width and speed follows the solar cycle. There is no distinct and markable difference in the properties of $CME_{SLD}$ associated with the prominence.

Table 8: List of the flares accompanied by $(CME)_{sld}$

| Cycle 23 | Cycle 24 |
|---|---|

| Sr. no. | Flare date (class) | Start time (HH:MM) UT | CME (HH:MM:SS) UT | CME width (Degree) | CME speed (km/s) | Sr. no. | Flare date (class) | Start time (HH:MM) UT | CME (HH:MM:SS) UT | CME width (Degree) | CME speed (km/s) |
|---|---|---|---|---|---|---|---|---|---|---|---|
| 1 | 11/10/2004(B1.1) | 20:11 | 20:30:06 | 20 | 408 | 1 | 26/06/2016 (B2.1) | 10:46 | 10:48:05 | 8 | 492 |
| 2 | 14/10/2006 (B1.3) | 02:14 | 03:30:06 | 33 | 217 | 2 | 26/06/2016 (B3.8) | 15:57 | 17:00:05 | 33 | 323 |
| 3 | 15/06/2007 (B2.0) | 04:35 | 05:06:04 | 12 | 297 | 3 | 27/06/2016 (B5.8) | 09:42 | 10:36:05 | 99 | 356 |
| 4 | 08/09/2007 (B3.0) | 00:13 | 01:31:41 | 34 | 98 | 4 | 28/10/2016 (B2.1) | 21:16:00 | 21:36:05 | 5 | 272 |
| 5 | 05/04/2008 (B1.3) | 01:47 | 02:26:04 | 11 | 166 | 5 | 04/01/2017 (B1.8) | 02:35 | 02:48:23 | 23 | 375 |
| 6 | 26/04/2008 (B3.8) | 13:54 | 14:30:04 | 281 | 515 | 6 | 10/01/2017 (B7.3) | 10:25 | 11:00:05 | 130 | 150 |
| | | | | | | 7 | 11/06/2017 (B1.5) | 01:07 | 02:00:05 | 93 | 304 |
| | | | | | | 8 | 21/11/2017 (B1.7) | 1:58 | 02:48:05 | 31 | 236 |
| | | | | | | 9 | 06/07/2019 (B3.2) | 21:16 | 21:48:05 | 52 | 110 |

Table 9: The summary of the properties of $CME_{SLD}$ the associated with prominence.

| Cycle 23 | | | | | | Cycle 24 | | | | | |
|---|---|---|---|---|---|---|---|---|---|---|---|
| Year | No. of event | Mean A.W. (degree) | Average Speed (km/s) | Average mass $\times 10^{14}$ (gm) | Average Kinetic energy $\times 10^{29}$ (erg) | Year | No. of event | Mean A.W. (degree) | Average Speed (km/s) | Average mass $\times 10^{14}$ (gm) | Average Kinetic energy $\times 10^{29}$ (erg) |
| 2005 | 4 | 50.75 | 391.75 | 1.03 | 0.69 | 2016 | 2 | 51 | 215.5 | 0.65 | 1.93 |
| 2006 | 10 | 40.1 | 330.9 | 14.5 | 13.9 | 2017 | 2 | 13.5 | 318 | | |
| 2007 | 17 | 20.05 | 238.82 | 3.52 | 1.83 | 2018 | 11 | 31.18 | 181.27 | 0.96 | 0.28 |
| 2008 | 16 | 20.5 | 290.18 | 6.6 | 2.8 | 2019 | 07 | 12.14 | 219.42 | | |

## 4. Summary and discussion:

The progression of increase in the SLD number has been observed from the solar cycle 21 to 24. In consequence, the cycle 24 shows the highest numbers of SLD as measured with the previous three cycles. The descending phase of cycle 23 and 24 noted 514 and 623 SLDs respectively, exhibits an increase of 21% (109). This continual increase in the SLDs from cycle 21 to 24 indicates, sequential weakening of magnetic field generation in the solar interior (Georgieva et al. 2017; Kakad et al. 2019). Similarly various researcher studied the different photospheric activity and solar magnetic field which evidence that the cycle 23 and 24 is likely to be the part of a series of 2-4 successive weak and long cycles may be the part of the "new grand minimum" (Yousef 2003; Feynman et al. 2011; Pevtsov et al. 2014). The number of SLD (Burud et al. 2021) and Magnetic helicity released (Hawkes et al. 2018) in the descending phase of a cycle can be a useful tool to prediction of solar activity in the next cycle. The helicity ejected by CMEs affect the differential rotation (DeVore 2000). In this connection we studied the $CME_{SLD}$ activity and its characteristics in detail. Further SOHO/LASCO provide a chance to analyzes and compare the properties of the $CME_{SLD}$. The result of comparative analysis summarizes as follows:

1) During the descending phase of the cycle 24, the SLD are increased by the 21% in contrast to that CME occurrence rate decreased by only 8.3%. One of the reasons for this discrepancy is the CME that occurred on SLD.

2) The number of SLD, $CME_{SLD}$ and average annual acceleration of $CME_{SLD}$ has anticorrelation with the sunspot activity while average annual linear speed and angular width are well correlated with solar cycle.

3) The SLD and $CME_{SLD}$ has a high correlation with the occurrence rate of $CME_{SLD}$ is 1.55 and 1.05 for cycle 23 and 24 respectively.

4) $CME_{SLD}$ events in cycle 24 has higher angular width and Kinetic energy but same speed and mass as that of the cycle 23.

5) The accelerated events are almost double in frequency compared to the decelerated event. It has shown anticorrelation with the solar cycle progression.

6) Very few numbers of solar flare, prominence were associated with $CME_{SLD}$. On the other hand, in rare cases $CME_{SLD}$ will produces minor geomagnetic storms.

The fraction of CMEs that occurred on SLDs; known as $CME_{SLD}$ is found to be 14 and 11%, for the descending phases cycles 23 and 24, respectively. Our analysis shows that the existence of sunspots is not a sufficient condition solar eruptive activity our result agrees with the Gopalswamy et al., (2010) and Vourlidas and Webb (2018). Hence solar emissions may have sources rather than the active regions. After the polar field reversal in year 2004, the strength of polar magnetic field and radial magnetic field has decreased (Petrie 2015; Wang 2009) which reduces heliosphere pressure. This may explain the observational result that CME in cycle 24 were wider and estimated to carry higher mass and Kinetic energy (Michalek et al.2019; Gopalswamy et al. 2015).

Further, we compare the characteristics such as speed, width, and acceleration of this sample size of $CME_{SLD}$ with the all CME that occurred in the descending phase by using the technique purposed by D'Huys et al.(2014). The analysis revealed that the $CME_{SLD}$ properties are, in general, do not show any clear distinction from the randomly selected events that occur on the non-SLDs. Thus, establishing the unique nature of the events, and rejecting a possibility of the selection bias. The average values of speed, width, and acceleration of the $CME_{SLD}$ always less than but do not have any distinct difference from that of the all-CME distribution expect minimum phase. This proves that the $CME_{SLD}$ are slower, smaller in width. The annual variation of the speed and width of the $CME_{SLD}$ follows the solar cycle; while the acceleration shows the anticorrelation. Our result conforms the result obtained by Du (2021); the CME acceleration depends more on the strength of solar activity not on the CME's speed. The annual average values of CME speeds and widths during SLD and non-SLD exhibited a declining trend in the descending phase of the solar cycle, a continual increase in the acceleration is spotted in general. This may possibly be associated with the effect of drag force exerted by the solar wind on slow CMEs (Kay 2022). We further found the intensity of the storm to be well correlated with the product V and Bz. Besides, although it is crucial to investigate the CME properties in the spotless days, the investigation over several solar cycles is deemed necessary to attain statistical significance for its prediction based on source properties (e.g., sunspot numbers, flares, etc.).

Data availability statement. Data can be downloaded at:

ftp://ftp.ngdc.noaa.gov/STP/swpc_products/weekly_reports/PRFs_of_SGD/

https://hinode.isee.nagoya-u.ac.jp/flare_catalogue/.

URL: https://cdaw.gsfc.nasa.gov/CME_list/.

http://voparis-helio.obspm.fr/hfc-gui/index.php

https://omniweb.gsfc.nasa.gov/form/dx1.html)

(https://wdc.kugi.kyoto-u.ac.jp/index.html)

Acknowledgements. Authors received the data from the Heliophysics Feature Catalogue (HFC), CDAW catalog, SGD report and Hinode flare catalogue. The LASCO CME catalog is generated and maintained at the CDAW Data Center by NASA and The Catholic University of America in cooperation with the Naval Research Laboratory. Cooperation received from President, KSV is sincerely acknowledged. A.K.A. is supported by the National Science Centre, Poland grant No. 2023/49/B/ST9/02409. **We earnestly thank the referee for valuable remarks which improved the paper.**